\documentclass[twocolumn]{revtex4}
\usepackage{graphicx}
\usepackage[]{epsfig}
\newcommand{\beq}{\begin{equation}}
\newcommand{\eeq}{\end{equation}}
\newcommand{\beqd}{\begin{displaymath}}
\newcommand{\eeqd}{\end{displaymath}}
\newcommand{\beqa}{\begin{eqnarray}}
\newcommand{\eeqa}{\end{eqnarray}}

\newcommand{\comment}[1]{}

\newcommand{\bew}{\begin{widetext}}
\newcommand{\eew}{\end{widetext}}

\begin{document}

\title{Solvable Models of Supercooled Liquids in Three Dimensions}
\author{Tommaso Rizzo}
\affiliation{Dip.\ Fisica, Universit\`a ``Sapienza'', Piazzale A.~Moro 2, I--00185, Rome, Italy}
\affiliation{ISC-CNR, UOS Rome, Universit\`a ``Sapienza'', Piazzale A.~Moro 2, I-00185, Rome, Italy}
\author{Thomas Voigtmann}
\affiliation{Deutsches Zentrum f\"ur Luft- und Raumfahrt (DLR), 51170 K\"oln, Germany}
\affiliation{Department of Physics, Heinrich-Heine-Universit\"at D\"usseldorf, Universit\"atsstra\ss{}e 1, 40225 D\"usseldorf, Germany}


\begin{abstract}
We introduce a supercooled liquid model  and obtain parameter-free quantitative predictions that are in excellent agreement  with numerical simulations, notably in the hard low-temperature region characterized by strong deviations from Mode-Coupling-Theory behavior. The model is the Fredrickson-Andersen Kinetically-Constrained-Model on the three-dimensional $M$-layer lattice. The agreement has implications beyond the specific model considered because the theory is potentially valid for many more systems, including realistic models and actual supercooled liquids.
\end{abstract}

\maketitle

The ubiquity of glass in nature and technology has driven research in this area for decades but there  is still no agreement on the basic mechanism by which a supercooled liquid forms a glass \cite{biroli2013perspective,wolynes2012structural,gotze2008complex}.
An essential difficulty for deciding over competing theories is the fact that often they make only {\it qualitative} statements. For instance, the classic thermodynamics vs. dynamics controversy resolves around the putative divergences of susceptibilities and correlation lengths but, while the debate has produced many conceptual developments \cite{biroli2013perspective,wolynes2012structural}, there are no precise {\it quantitative} predictions to be be matched with experiments and numerical simulations. 
In other words, models of supercooled liquids in three dimensions, either realistic or on lattice, have been studied previously only by means of numerical simulations. In the following we present a model in three dimensions in which the gap between theory and experiments can instead be filled. 
Furthermore the agreement has implications beyond the specific model considered because the theory is potentially valid for many more physical systems, including the most realistic models and actual supercooled liquids.

Mode-Coupling-Theory (MCT) \cite{gotze2008complex} is widely popular in the experimental literature because it captures many {\it qualitative} features of the physics of liquids upon supercooling, notably two-step relaxation and stretched exponential decay.
Furthermore it  agrees {\it quantitatively} with numerical simulations 
although one has to replace the values of some MCT parameters with values extracted from data fits \cite{nauroth1997quantitative,kob1999computer,sciortino2001debye,weysser2010structural}. 
The essential problem is that it predicts a dynamical arrest transition
at a temperature  where actual systems are still in the liquid phase.
In spite of this serious drawback many believe that the MCT transition is still relevant and determines a crossover from power-law to exponential increase of the relaxation time that is widely observed.
Further support to this scenario comes from the fact that simple liquid models in the limit of infinite physical dimension $d$ display a sharp transition  qualitatively similar to the one of MCT, although MCT itself is quantitatively wrong in this limit \cite{charbonneau2017glass}. In that case the sharp transition is clearly a mean-field artifact due to the $d\rightarrow \infty$ limit and it should become a crossover as soon as the dimension is finite.  
Similarly, the success of kinetically constrained models (KCM) in reproducing the physics of supercooled liquids is often attributed to the presence of an avoided MCT-like transition that becomes sharp when one switches from lattices in finite dimensions to the Bethe lattice where mean-field theory is correct  \cite{fredrickson1984kinetic,fredrickson1985facilitated,sellitto2005facilitated,sellitto2015,de2016scaling,
franz2013finite,ikeda2017fredrickson,sausset2010bootstrap}.

Building on the analogy between MCT and spin-glass models with one step of Parisi's Replica-Symmetry-Breaking (1RSB) discovered more that thirty years ago \cite{kirkpatrick1987p}, it has been recently proposed that to fix MCT one has to replace it with a set of  stochastic dynamical equations called Stochastic-Beta-Relaxation (SBR) \cite{rizzo2014long,rizzo2016dynamical}. 
SBR describes the $\beta$ regime, {\it i.e.} the time-scale when dynamic correlations stay close to a plateau value, and it has a simple and intuitive interpretation: it is basically MCT supplemented with random spatial fluctuations of the temperature that are effectively quenched on the $\beta$ time-scale. 
SBR is promising as it seems to cure the drawbacks of MCT, displaying in particular the crossover and also dynamical heterogeneities, without spoiling its successes, in particular two-step relaxation and stretched exponential decay \cite{rizzo2015qualitative,rizzo2015nature}.
SBR is a universal theory potentially valid for many different microscopic models, the specific model determining the values of its few (five) quantitative parameters.  In the following we show that SBR holds for the models we study. To do so we have first computed the five SBR parameters, then solved numerically the dynamical stochastic equations and finally performed  Monte-Carlo simulations to be compared with the theoretical predictions. It turned out that SBR provides an accurate quantitative parameter-free description of the dynamics, {\it i.e.} the models are {\it solvable} beyond mean-field theory.

\begin{figure}
\centering
\includegraphics[scale=.5]{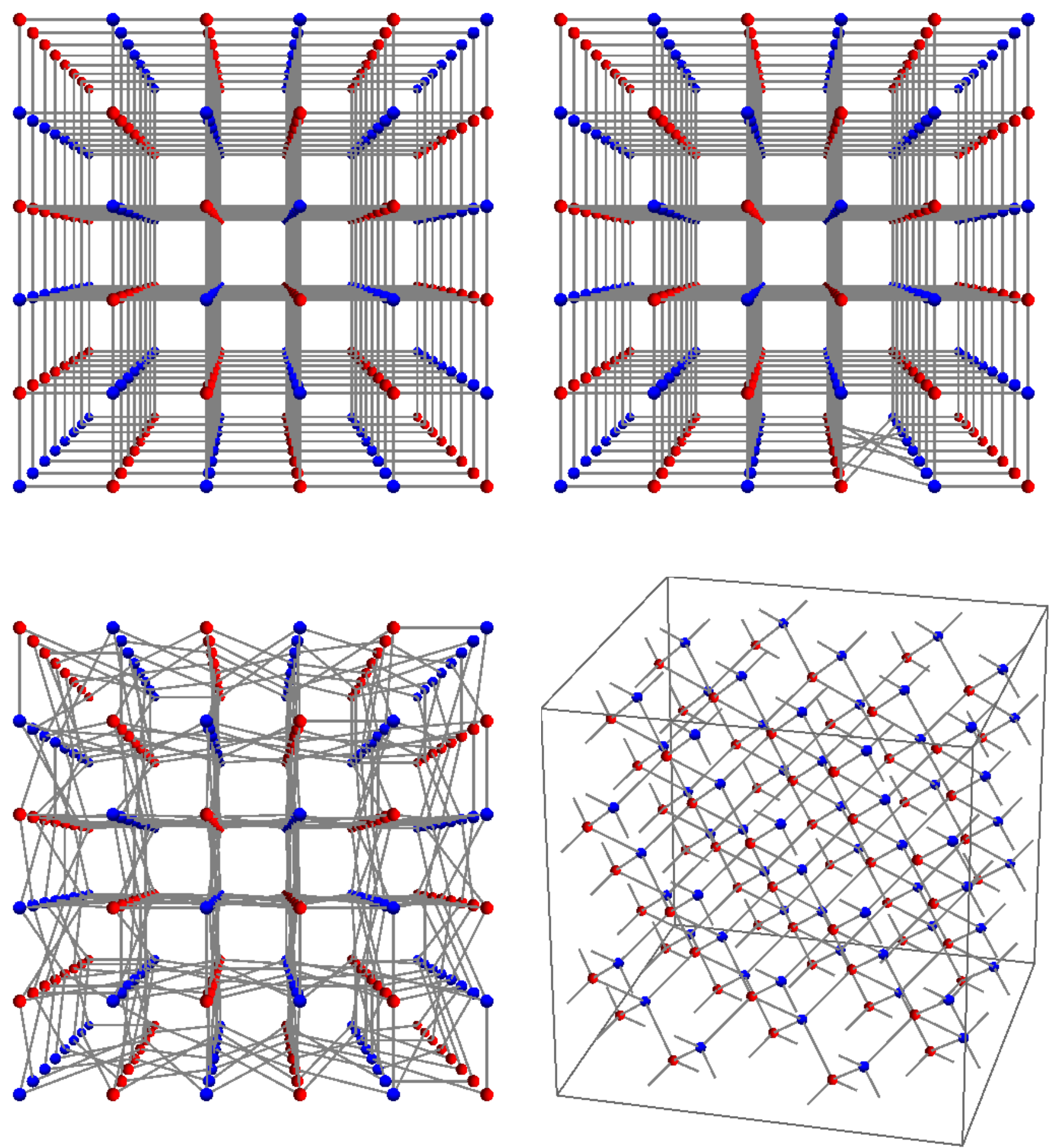} 
 \caption{left to right, top to bottom: The $M$-layer procedure: 1) $M$ independent copies of a lattices (a regular 2D lattice in the figure) are stacked on top of each other (view from the top) 2) the edges corresponding to a given link in the original lattice are rewired between the $M$ copies, 3) the procedure is repeated for every link of the original lattice. For large $M$ the graph is locally a Bethe lattice but at large distances it has the properties of finite-dimensional lattice. 4) The three dimensional diamond cubic lattice}
\label{fig:ML2d}
\end{figure}

The model considered is a particular realization of the classic Fredrickson-Andersen (FA) KCM \cite{fredrickson1984kinetic,fredrickson1985facilitated}.
The FA model is made by Ising spins on the sites of a lattice that are independent, the Hamiltonian being 
 $H=\sum_i s_i$,  but obey a kinetically constrained dynamics: a spin can flip only if it has at least $m$ of its $c$ nearest neighbors in the excited (up) state. 
An equilibrium configuration is thus easily generated numerically at time zero and one typically measures the persistence. More precisely, we define the local persistence $\phi_i(t)$ as equal to one if $s_i(t')=-1$ for all $0 \leq t' \leq t$ and zero otherwise, thus the averaged persistence is  the number of {\it negative} sites that have never flipped at time $t$ divided by the total number of spins \footnote{Note that, at variance with ours, the definition of persistence often used in the literature takes into account all sites that did not flip, both negative and positive. The quantitative agreement between numerical data and theory is expected to hold also using the conventional definition of the persistence or other proxies of the correlation, {\it e.g.} the overlap. This can be understood within the Renormalization Group framework to be discussed later in the text as all observables have the same behavior provided they have a non-zero projection on the critical mode. Quantitatively, different (shifted) observables are just related by a rescaling factor. 
The rationale for our choice is that analytic computations of the quantitative SBR parameters, see discussion later and details in the supplemental material, are simpler because our definition is as close as possible to bootstrap percolation.}. 
The FA model on the Bethe lattice is known to exhibit a dynamical arrest transition of the MCT type \cite{sellitto2005facilitated,sellitto2015,de2016scaling,
franz2013finite,ikeda2017fredrickson,sausset2010bootstrap}:
at the critical temperature $T_c$ the  persistence remains blocked to a plateau value $\phi_{plat}$ that is approached in a power-law fashion. The FA dynamical transition is intimately related to bootstrap percolation (BP) and both $T_c$ and $\phi_{plat}$ can be computed from its solution on the Bethe lattice as discussed in the supplemental material.
In particular for connectivity $c=4$ and $m=2$ the average persistence $\phi(t)$ obeys at $T_c=0.480898$:
\beq
\phi(t)- \phi_{plat}\, \approx {1 \over (t/t_0)^a}\,   , \ t \gg1 \, ,
\label{Betheg}
\eeq
where $\phi_{plat}=21/32$.
At present, analytic expressions of $t_0$ and $a$ are not available but they can be estimated  from numerical simulations as $a \approx 0.352$ and $t_0 \approx 2.30$.
It is well-known that the sharp transition is instead {\it avoided} when the FA model is studied on regular lattices in two and three dimensions but no first-principle theoretical description of the dynamics can be obtained in those cases. 
As we will show in the following, such a description can instead be obtained on the finite dimensional lattice we consider here.

We studied the FA model (with $m=2$) on the (random) lattice in three dimensions yielded by the application of $M$-layer construction of \cite{altieri2017loop} to the diamond cubic lattice (that has connectivity $c=4$, see fig. \ref{fig:ML2d}).
The $M$-layer construction can be applied to any lattice: to obtain a random instance
one considers $M$ copies of the original lattice, say the square lattice in $d=2$ as in the figure, rewires through a random permutation the $M$ links corresponding to a given link on the original $(M=1)$ lattice and finally repeats the procedure for each link of the original lattice as shown in Fig. (\ref{fig:ML2d}).  
It can be easily seen that short loops in the lattice are rare for large values of $M$ and the lattice is locally tree-like. Loops are nevertheless present at large distances, thus at any finite $M$ the lattice is finite dimensional although it looks like a Bethe lattice at short distances. 
Given that  for each site $i=1,\dots, N$ of the original lattice ($M=1$)  there are $M$ spins $s_i^\alpha$, $\alpha=1,\dots,M$  the total number of sites is $N_{tot}=M \times N$ and
the natural local order parameter is the average over the layers of the local persistence minus the plateau value
\beq
g(x,t) \equiv \left({1 \over M} \sum_{\alpha=1}^M  \phi_i^\alpha (t)\right)- \phi_{plat}\, ,
\eeq
where $x$ is the spatial coordinate of site $i$.
For $M$  large but finite it is natural to expect that any observable takes the same value it has on the Bethe lattice with small $O(1/M)$ corrections: the model should be solvable by the Bethe approximation.
This is indeed true {\it except} at the Bethe lattice critical temperature: while the Bethe approximation predicts that the averaged order parameter $g(t)$ never reaches zero the three-dimensional nature of the $M$-layer lattice implies that this cannot be true and leads to a dramatic deviation from mean-field behavior.
\begin{figure}[t]
\centering
\includegraphics[scale=.7]{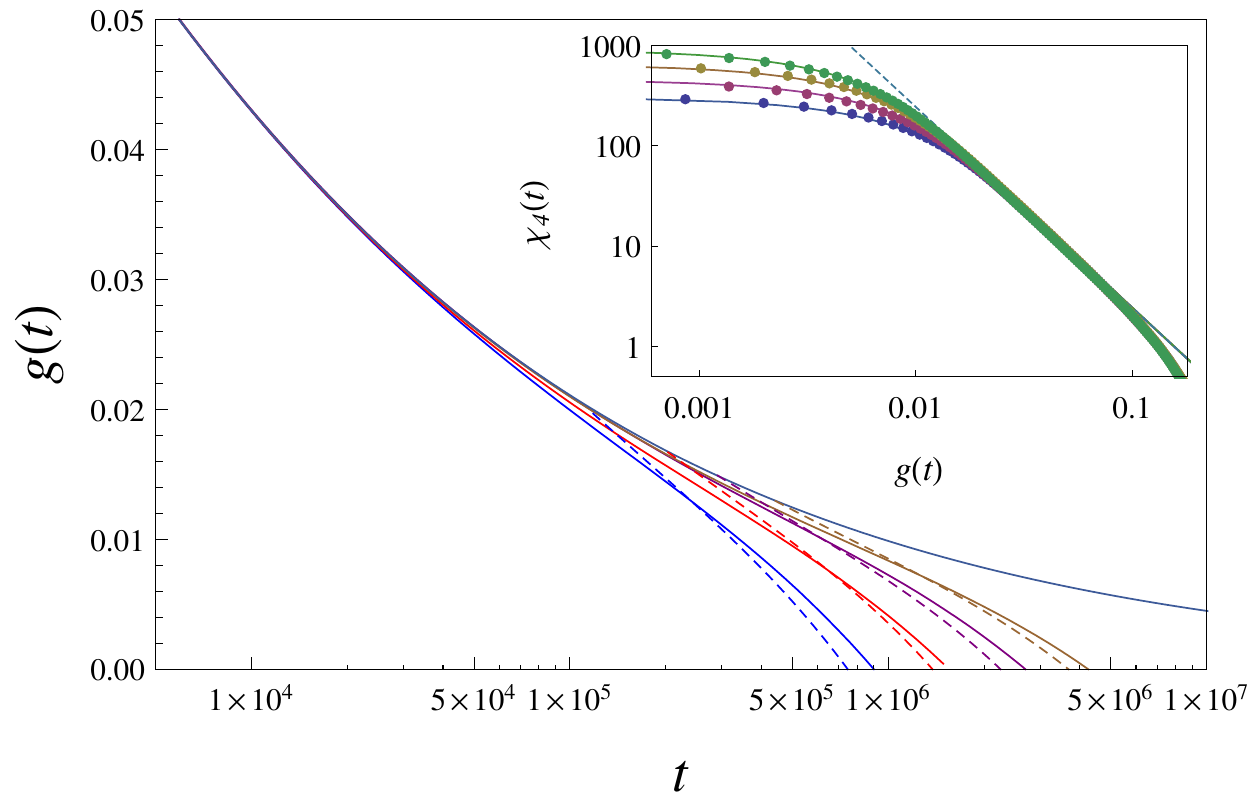} 
 \caption{Persistence vs. Time in $d=3$. Solid lines from bottom to top: data for $M=3000$,  $M=50000$, $M=100000$, $M=200000$ and for the Bethe lattice curve ($M=\infty$). The data follow the Bethe curve at initial times  and deviate from it at later times increasing with $M$. The dotted lines represent the corresponding SBR predictions describing the data when they start to deviate from the mean-field curve, see text.
 Inset: Parametric plot of $\chi_4(t)$ vs. $g(t)$ in $d=3$ for the above values of $M$. The SBR predictions (solid lines) are in excellent agreement with the numerical data (points) in the large time regime where $g(t)$ deviates from MF. At small intermediate times (large $g(t)$) both SBR and the numerical data approaches the MF asymptote (dashed straight line), numerical data deviates on smaller (microscopic) times.}
\label{fig:pp3D}
\end{figure}
This is indeed confirmed by numerical data. In fig. (\ref{fig:pp3D}) we plot the time decay of the shifted persistence  $\langle g(x,t)\rangle$ where here and in the following the angle brackets  mean average with respect to: i) different instances of the random lattice generated by the $M$-layer construction, ii) different initial equilibrium configurations and iii)  different thermal histories.
Monte Carlo simulations were carried  at the critical temperature of the Bethe lattice $T_c=0.480898$  for systems with $M=3000$ (with size of the original diamond cubic lattice $L=8$),  $M=50000$, $M=100000$, $M=200000$ ($L=4$). At initial times the data follow the mean-field (MF) curve corresponding to $M=\infty$ (obtained from numerical simulations on the Bethe lattice (see the SI) and deviate from it at larger times (increasing with $M$) reaching the plateau value ($g=0$) in finite time.

The difficult problem is to obtain theoretical predictions in the region where the data deviate from the mean-field curve and to proceed it is useful to examine {\it the role of fluctuations}.
Within the $M$-layer a mean-field approximation to fluctuations can be obtained as a sum over non-backtracking walks of Bethe lattice fluctuations \cite{altieri2017loop}.
At the critical temperature this yields  \footnote{See details in the Supplemental Material, which includes Refs. \cite{parisi2013critical,schwarz2006onset,franz2011field,fitzner2013non}}:
\beq
\langle g(x,t) g(y,t) \rangle- \langle g(x,t) \rangle \langle g(y,t) \rangle \stackrel{\text{MF}}\approx  {t^{a (2 -d /2)} \over M}  f\left( {x-y \over \xi(t) }\right)\ .
\label{MFprop}
\eeq
where $f(x)$ is a scaling function, the correlation length diverges with time as $\xi(t) \propto t^{a/2}$ ansd $d$ is the space dimension. We can now invoke the {\it Ginzburg criterion} and argue that the MF approximation $\langle g(x,t) \rangle \stackrel{\text{MF}}\approx 1/(t/t_0)^a$ is accurate as long as fluctuations around the mean are small.
Generically they are small due to the $1/M$ prefactor but we see  that there is a time-scale $t_G$ when, due to the $t^{a (2 -d /2)}$ prefactor they become comparable with the (squared) order parameter $\langle g(x,t) \rangle \stackrel{\text{MF}}\approx 1/(t/t_0)^a$:
\beq
 {t_G^{a (2 -d /2)} \over M} \approx {1 \over t_G^{2a}} \rightarrow t_G \approx M^{{1 \over a( 4-d/2)}} \ .
\label{Ginzburgtime}
\eeq 
Thus on this time-scale MF theory must fail and most notably the spurious transition will be avoided, for instance the dressed propagator on the LHS of (\ref{MFprop}) will deviate from the bare expression on the RHS and the actual correlation length will cease to grow.
We note that the Ginzburg time grows with $M$ and thus deviations from MF occur at later time for increasing values of $M$ in agreement with the data of fig. (\ref{fig:pp3D}).
Most importantly  since $t_G \approx M^{1 \over a( 4-d/2)}$ is large for large $M$, the order parameter  
is $O(1/t_G^a)$ small and the correlation length is $O(t_G^{a/2})$ large: this grants that deviations from mean-field theory are described by an effective Landau theory, because one can retain only the lowest orders in the Taylor expansion of the order parameter and its space and time derivatives. 
Following the arguments and computations of \cite{rizzo2014long,rizzo2016dynamical} we argue that the effective theory is SBR, meaning that the generic $K$-point average obeys for $1\ll M < \infty$:
\beq
\langle g(x_1,t_1) \dots g(x_K,t_K) \rangle \approx [\hat{g}(x_1,t_1) \dots \hat{g}(x_K,t_K)]\, .
\label{equivalence}
\eeq
where  $\hat{g}(x,t)$ in the RHS is the solution of the SBR equations:
\beq
\sigma  + s(x) =- \alpha\, \nabla^2 \, \hat{g}(x,t)-\lambda \, \hat{g}^2(x,t)+{d \over dt}\int_0^t \hat{g}(x,t-s)\hat{g}(x,s)ds \ .
\label{SBRequa}
\eeq
The separation parameter $\sigma$ measures the distance from the critical point and vanishes at $T=T_c$. The square brackets mean average with respect to  the field $s(x)$ that is a quenched random fluctuation of $\sigma$, Gaussian and delta-correlated in space:  
\beq
[s(x)]=0\, ,\ [s(x)s(y)]=  \Delta \sigma^2 \, \delta (x-y) \ .
\eeq
the SBR equations have to be solved with the small-time condition
$\lim_{t \rightarrow 0} \hat{g}(x,t) (t/t_0)^a=1$ 
where $\lambda$ and $a$ are related by the MCT relationship $\lambda={\Gamma^2(1-a) \over \Gamma(1-2a)}$.
In practice for times smaller than $t_G$ the observables on the LHS of eq. (\ref{equivalence}) can be accurately approximated with the values they have on the Bethe lattice while on times of order $t_G$ they are described by the RHS. This explains the peculiar initial conditions of the SBR equations: the {\it short-time} behavior on times $O(t_G)$ matches {\it long-time} behavior for times  $1 \ll t \ll t_G$, {\it i.e.} the mean-field result given by eq. {\ref{Betheg}}.
  
Equation (\ref{equivalence}) embodies the power of the effective theory approach: on the LHS we have a model with a complex microscopic dynamics for which no analytic treatment of dynamics is available (not even on the Bethe lattice), on the RHS we have a (numerically) solvable set of equations  that were derived in \cite{rizzo2014long,rizzo2016dynamical} starting from symmetry considerations (essentially the detailed balance property of the dynamics) but {\it without} reference to any specific microscopic model.
The microscopic details determine the actual values of the five SBR parameters $a$, $t_0$, $\alpha$, $\Delta \sigma$ and $\sigma$ that  are needed to get quantitative predictions.
Using recent developments on bootstrap percolation on the Bethe lattice \cite{rizzo2019fate} and some lattice-dependent geometrical constants, we obtain (details in the SI):
\beq
\Delta \sigma^2={0.285 \over M } \,\, ,\ \alpha=0.411 \, , \ \sigma=0.222 \times (T_c-T)\ . 
\label{barecouplings}
\eeq
Within SBR, mean-field theory is recovered setting $\Delta \sigma^2=0$, in this case $\hat{g}(x,t)$ is constant in space, the gradient term plays no role and one recovers the critical MCT equation \cite{gotze2008complex}, in particular for $\sigma \geq 0$ ($T<T_c$) $\phi(t)$ never goes below the plateau value. The $M$-layer construction allows to have a finite but small $\Delta \sigma$ so that the MCT transition is avoided and  $\phi(t)$ crosses the plateau at a finite time for all values of $\sigma$. The SBR predictions corresponding to the data shown in fig.  (\ref{fig:pp3D})  were obtained solving numerically (by space-time discretization) eq. (\ref{SBRequa}) for many instances of the $s(x)$ in a box of size $L$. 
From the figure we note that the quality of the SBR predictions increases with $M$ and is excellent for $M=200000$, especially considering that there is {\it no single fitting parameter}.

SBR is a powerful theory that provides not only the average dynamical order parameter but, according to eq. (\ref{equivalence}), also {\it all possible fluctuations}. To demonstrate this, in the inset of  fig. (\ref{fig:pp3D}), we plot parametrically the $\chi_4(t)$ function that yields the fluctuations of the persistence density:
\beq
g(t) \equiv {1 \over N_{tot}}\sum_{i,\alpha} g_i^\alpha(t)\, ,\ 
\chi_4(t) \equiv N_{tot}\,(\langle g^2(t)\rangle-\langle g(t)\rangle^2)\ .
\eeq
According to the MF expression (\ref{MFprop}) $\chi_4(t)$ should diverge with time as $t^{2 a}$, (leading to a MF asymptote $\chi_4(g) \propto g^{-2}$), instead on the Ginzburg time scale $t_G$ over which $g(t)$ deviates from MF and reaches zero $\chi_4(g) $ deviates from the MF  law and remains finite. Note that the agreement between numerical data and the numerical solution of the SBR equations is even better in the parametric representation.

SBR can be applied in other dimensions as well, we have considered (SI) the $d=0$ case  that corresponds to finite-size effects in mean-field models on fully-connected or sparse random graphs with 1RSB at the so-called dynamical temperature $T_d$ \cite{mezard2009information}. The interplay between the parameters $M$ and $L$  can be also be clarified in terms of the SBR equations (SI). 

To discuss the results in a broader context we note that the $M$-layer construction can be applied  virtually to all supercooled liquids lattice models, including different KCM's \cite{ritort2003glassy} and plaquette models \cite{jack2016phase,biroli2016role}, leading to analogous solvable non-MF models described by SBR. Furthermore SBR can provide quantitative  theoretical predictions for generic tunable models \cite{franz2004kac,franz2007analytic,caltagirone2011ising,
mari2009jamming,mari2011dynamical,charbonneau2014hopping,
berthier2012finite}  that in earlier studies could only be studied by means of numerical simulations.  
On the other hand, the lattice for $M \gg 1$ is rather different from the original $M=1$ lattice one is ideally interested in: the latter has many short loops while the former has very few. Thus the condition $M \gg 1$  alters artificially the three-dimensional geometry at the microscopic scale and one may ask if this hampers the applicability of SBR to realistic models and actual supercooled liquids.
To clarify this point we stress that $M \gg 1$ is a sufficient but {\it not} necessary condition. A {\it necessary} condition in a generic system is that the dynamical  correlation length is large enough to justify the use of a coarse-grained description: numerical simulations do indeed report correlation lengths significantly larger than the microscopic scale in supercooled liquids \cite{lavcevic2003spatially,berthier2007spontaneous,
flenner2013dynamic,karmakar2014growing} while unfortunately they cannot be measured in current experimental settings.
The natural framework to discuss coarse-grained observables is Wilson's renormalization-group (RG) theory where each system corresponds  to  a particular point in the space of RG Hamiltonians that display all possible powers of the  order parameter and its spatial and time derivatives and thus depends on a infinite number of coupling constants. In practice these additional terms lead to higher powers of $\hat{g}(x,t)$ and higher orders spatial and time derivatives in eqs. (\ref{SBRequa}).
SBR assumes that these additional terms can be neglected and this can be motivated by the following RG argument. The absence of a sharp dynamical transition in finite dimension implies the absence of a stable fixed point (FP), as a consequence all Hamiltonians flow under RG towards the high-temperature FP, {\it howhever}, {\it if} the correlation length is large it will take many RG steps for it to decrease to one.
Since standard dimensional analysis implies that the coupling constants of the additional terms decrease close to the Gaussian fixed point (they are irrelevant operators in RG jargon) it is possible that on the scale of the correlation length the coarse-grained theory is driven near SBR by the RG flow.
This explain why  many different systems, including experimental ones, are potentially described by SBR and thus share the same {\it qualitative} features that does not depend on the actual values of the SBR parameters: notably power-law to exponential increase of the $\beta$ time and dynamical heterogeneities below the avoided transition.

SBR is thus potentially valid for $M=1$ as well, because long-range correlations may develop also in presence of short-range interactions. The deviations at smaller $M$ in the plots are indeed expected because the SBR parameters, being model-dependent, {\it change} with $M$.  The values we computed  in eqs. (\ref{barecouplings}) have actually $1/M$ corrections that can be also computed systematically through a feasible but tedious power expansion.
The only special feature of the large-$M$ regime is that the SBR parameters can be computed exactly from the Bethe lattice while the computation in the $M=1$ case is less straightforward (one should take into account the presence of small loops) but it is still feasible in principle.

While it is satisfying to compute the SBR parameter independently as we have done here, one could also extract some or all of them from fits. This means that SBR can be a useful tool to rationalize experimental data in the region where the widely used ideal MCT scalings fail.
The outcome would still be highly non trivial because the SBR eqs.  (\ref{SBRequa}) yields predictions for many more quantities  than those needed to determine the SBR parameters through fits.
In particular here we considered only one temperature, but one may consider a whole range of temperatures \cite{rizzo2015qualitative} (corresponding to different values of $\sigma$ in eqs. (\ref{SBRequa})) and also study spatial correlations \cite{rizzo2015qualitative} and finite-size effects.

\begin{acknowledgments}
We acknowledge the financial support of the Simons Foundation (Grant No. 454949, Giorgio Parisi). 
\end{acknowledgments}

\bibliography{biblio.bib}

\clearpage

\widetext
\begin{center}
\textbf{\large Supplemental Materials: Solvable Models of Supercooled Liquids}
\end{center}

\tableofcontents

\section{Numerical Simulations}

All simulations were performed at the critical temperature of the FA Model with $m=2$ on the Bethe lattice with connectivity $c=4$.
The corresponding value can be obtained from the solution of Bootstrap percolation on the Bethe lattice given below.
We considered the random lattice obtained by the application of the $M$-layer construction to the diamond cubic lattice with periodic boundary conditions. The diamond lattice is generated by repeating in the three directions a basic unit cell of length $L=4$. In each unit cell there are eight lattice points that can be divided into two groups: blue lattice points have coordinates $(0,0,0)$, $(2,0,2)$, $(0,2,2)$ and  $(2,2,0)$, red lattice points have coordinates $(3,3,3)$, $(3,1,1)$, $(1,3,1)$ and  $(1,1,3)$. Each red (blue) lattice point is connected with its four  blue (red) nearest neighbors.  
We have used a Chessboard/Metropolis setting:  all red spins are updated sequentially and then all the blue spins are updated sequentially. 
The single spin update is made with a Metropolis move: a negative mobile spin is flipped with probability $e^{-\beta}$ and a positive mobile spin is flipped with probability one.
The chessboard setting is more convenient than the standard Random-site/Metropolis setting typically considered in the FA model as will be discussed below.

\section{Mean-Field Behavior}
\label{MFbehavior}

\begin{figure}[b]
\centering
\includegraphics[scale=.7]{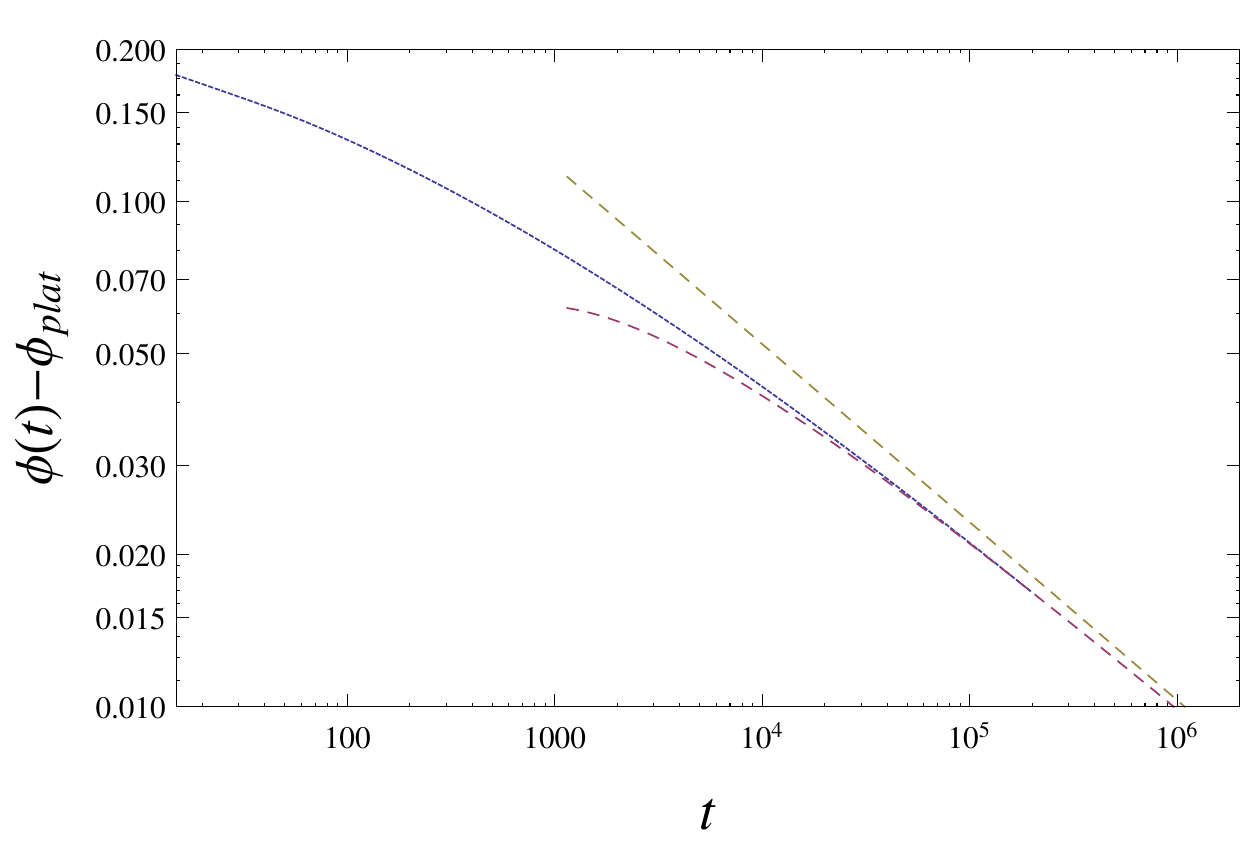}
 \caption{Shifted persistence at $L=4$ and $M=1.6 \times 10^6$ (solid), the dashed lines represent the asymptotic expression $(t/t_0)^{-a}$ (top) and the asymptotic  expression plus the negative sub-leading correction $c_1\,(t/t_0)^{-2a}$ (bottom).}
\label{fig:ptbethe}
\end{figure}

In figure (\ref{fig:ptbethe}) we plot $g(t)=\phi(t)-\phi_{plat}$ for a system with $L=4$ and $M=1.6 \times 10^6$.
In the range of times shown these data have converged on the Bethe lattice solution corresponding to $M \rightarrow \infty$.
\begin{figure}[b]
\centering
\includegraphics[scale=.7]{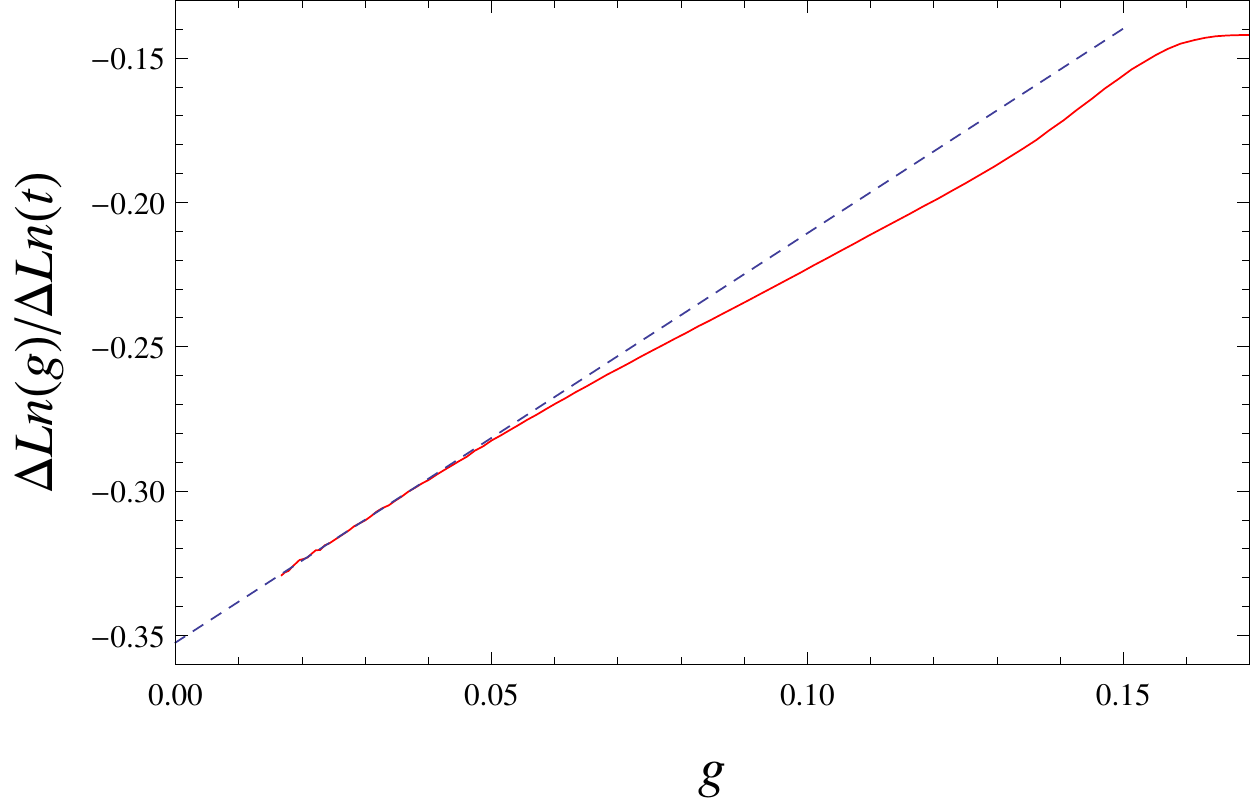} 
 \caption{Parametric plot of the (discrete) logarithmic derivative of $g(t)$ vs. $g(t)$ for  $L=4$ and $M=1.6 \times 10^6$. The dotted line is a linear fit $-a - a\, c_1 g$.}
\label{fig:dlog16E5}
\end{figure}
In fig. (\ref{fig:dlog16E5}) we plot parametrically the (discrete) logarithmic derivative of $g(t)$ vs. $g(t)$. For a $g(t)$ decaying at large times as $1/t^a$ the logarithmic derivative should converge to $-a$ in the limit of $g(t)\rightarrow 0$. The data display linear behavior in the small $g$ region corresponding to a correction of order $1/t^{2a}$. This is precisely the small-time correction that one would expect within MCT: the leading term is given by a quadratic equation while various subleading cubic terms induce a $1/t^{2a}$ correction.
In full generality we can write
\beq
g(t)= {1 \over (t/t_0)^a}+{\delta_1 \over (t/t_0)^{2a}}+\dots
\label{asymptotic}
\eeq
where all possible constants in front of the leading term are absorbed in the definition of $t_0$. The above expression leads to:
\beq
{\Delta \ln\, g \over \Delta \, \ln \, t}=-a - a\, \delta_1 g+ \dots 
\eeq
and thus from a linear fit we can extract $a$ and $\delta_1$.
The constant $t_0$ can then be estimated fitting the numerical data with the asymptotic form (\ref{asymptotic}), in fig. (\ref{fig:ptbethe}) we show both the $1/(t/t_0)^a$ term and the corrected expression.
We thus obtain the values quoted in the main text:
\beq
a \approx 0.352\, , \ t_0 \approx 2.30\ .
\eeq
From this we also have $\lambda=0.634$.
We stress that in the range of times accessed numerically the $1/t^{2 a}$ correction is significant and a linear fit of the data in fig. (\ref{fig:ptbethe}) would give an incorrect smaller exponent $a$. Instead the analysis of the second derivative is much safer and inconsistent with the value $a=.28-.3$ reported in earlier studies \cite{sellitto2005facilitated,sellitto2015,de2016scaling}.
We also note that  within the replica treatment the parameter exponent (and thus the exponent $a$) is available from the statics \cite{caltagirone2012critical,parisi2013critical} and only the coefficient $t_0$ must be extracted numerically. Unfortunately a static replica treatment of the FA model is not available at present.

\subsection{Comparison between different dynamics}

\begin{figure}
\centering
\includegraphics[scale=.7]{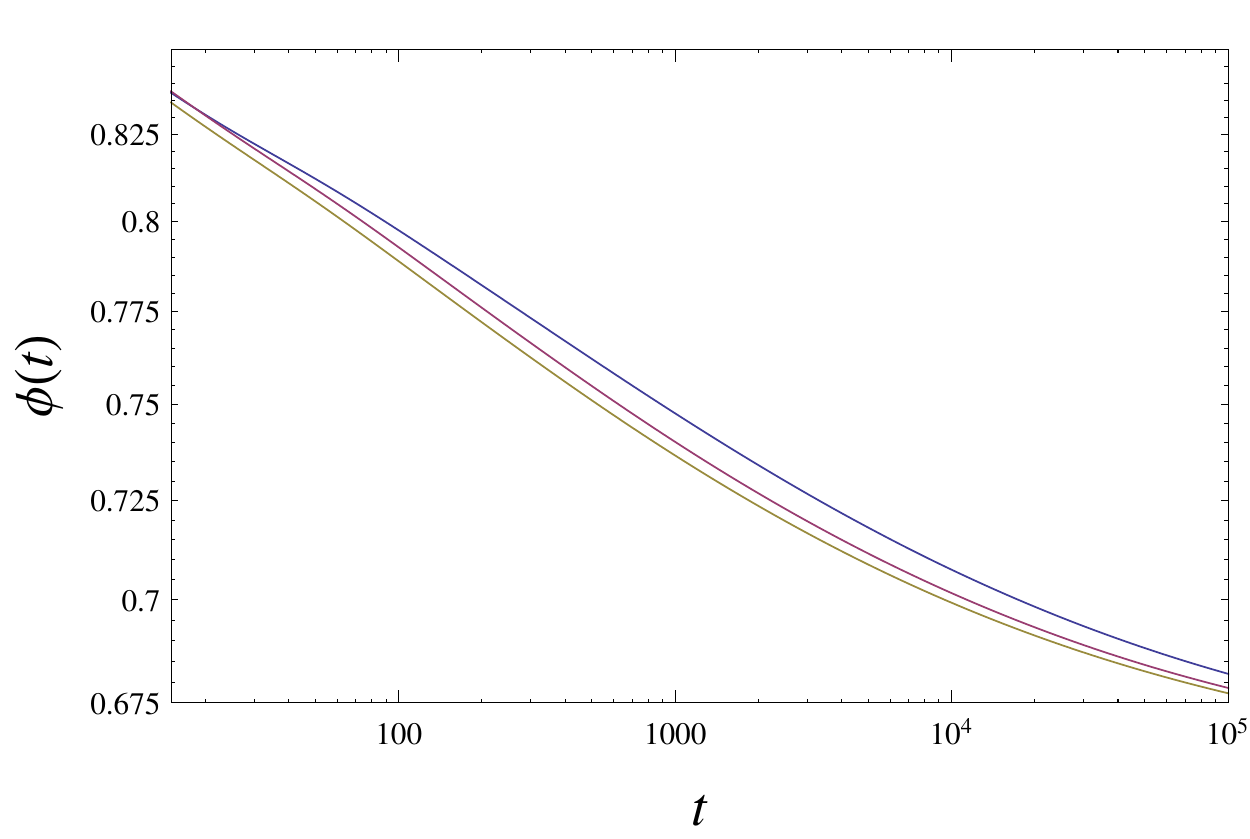} 
 \caption{Relaxation vs. time for a system with $L=4$ and $M=4\times 10^5$ with different dynamics, from top to bottom: Random-site/Metropolis, Chessboard/Heat-Bath, Chessboard/Metropolis}
\label{fig:ptdd}
\end{figure}
The choice of the Chessboard setting is more convenient than the Random-Site/Metropolis dynamics typically used in the literature because each Monte-Carlo step (MCS) requires less CPU time, besides in MCS unit the relaxation is faster as can be seen in fig. (\ref{fig:ptdd}) where we display the relaxation for a system with $L=4$ and $M=4\times 10^5$ with different dynamics, including Chessboard with heat-bath (Glauber) update. The key point is that {\it at large enough times the different curves differ only by a constant shift in time}: this can be seen more clearly considering the parametric plot of the logarithmic derivative of $g$ vs. $g$.
\begin{figure}
\centering
\includegraphics[scale=.7]{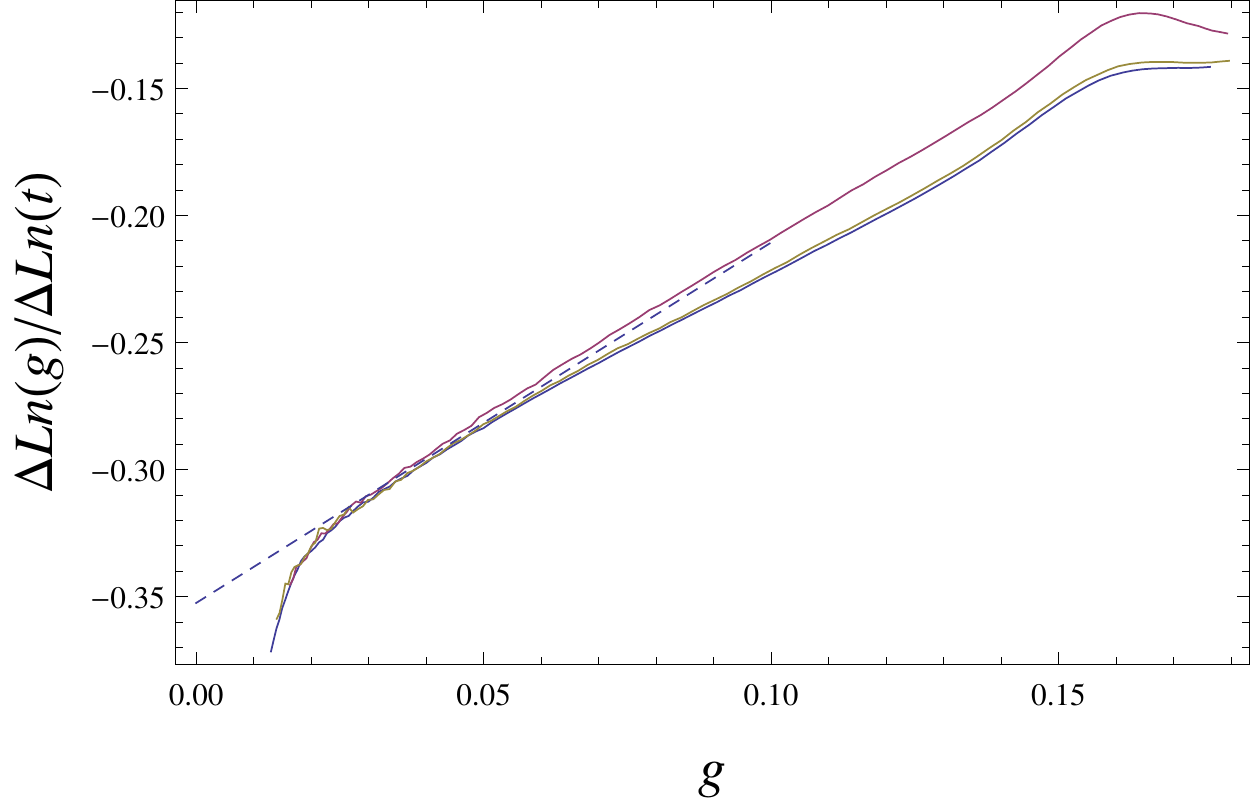} 
 \caption{Parametric plot of the (discrete) logarithmic derivative of $g(t)$ vs. $g(t)$ for  $L=4$ and $M=4 \times 10^5$, from top to bottom:  Random-site/Metropolis, Chessboard/Heat-Bath, Chessboard/Metropolis. The curves from different dynamics are on top of one another at small values of $g$. The dotted line is the linear fit describing the data on the Bethe lattice.}
\label{fig:dlog4E5}
\end{figure} 
In fig. (\ref{fig:dlog4E5}) we see that at small values of $g$ the parametric curves collapse onto a single curve independently of the dynamics. Note that the collapsing curve deviates from the linear fit corresponding to the asymptotic behavior implying that finite-size critical deviations are also independent on the dynamics. This is consistent with the fact that as we saw already these deviations are described by SBR in which the different microscopic dynamics enter only through the parameter $t_0$. We also mention that the Chessboard dynamic satisfies detailed balance but is not invertible, furthermore irreducibility of the Markov chain is not granted in general in sequential update with Metropolis moves, but these features, as usual, appear to be harmless in the interesting region of long-time critical behavior.

\section{Bootstrap Percolation on the Bethe lattice}
\label{sec:conn}

The dynamics transition of FA models on the Bethe lattice of connectivity $c$ is intimately related with $k$-core, or bootstrap, percolation \cite{sellitto2005facilitated}.
Let us recall the definition of BP: first the sites of a given
lattice are populated with probability $p$, then each site
with less than $k$ neighbors is removed and the culling process is repeated until each site has at least $k$ neighbors.
When the process is completed the remaining occupied sites, if any, form the so-called the $k$-core. 

One can argue that the cluster of blocked negative sites on the FA model is exactly the $k$-core with $k$ given by 
\beq
k=c-m+1
\eeq
therefore $k=3$ for the $(c=4,m=2)$ FA models considered here.
The relationship between temperature in the FA model and $p$ in BP is easily found:
\beq
p={e^\beta \over 1+ e^\beta}\ .
\eeq

The outcome of the culling process leading to the $k$-core does not depend on the sequence of which  sites are culled, in particular one can consider the  modified problem in which a given  site $i$ is occupied with probability one and no culling is applied to it and then  apply an {\it extraction and culling} (eac) move to obtain its probability $P_{site}$ to be in the $k$-core. $P_{site}$ is given by $p$ times the sum over all configuration of its neighbors $\{s_j, j \in \partial i\}$ satisfying the constraint of their occupancy probability  $P^{(i)}(s_j, j \in \partial i)$ {\it before} eac on site $i$.
On the Bethe lattice with connectivity $c$ one can argue that the probability $P^{(i)}(s_j, j \in \partial i)$ is factorized: 
\beq
P^{(i)}(\{s_j \, :\,  j \in \partial i\})= \prod_{j \in \partial i}P_B(s_j)\, \ ,
\eeq
where $s=-1$ if the site is occupied and $s=1$ otherwise and
\beq
P_B(s) \equiv \delta_{s,-1}\, P  + \delta_{s,1} (1-P) \ .
\eeq
With the definition $P_{s,t} \equiv \sum_{i=s}^{t}  {t \choose i} P^i (1-P)^{t-i}$ the Bethe solution is determined by the following equations:
\beq
P=p \, P_{k-1,c-1} \, , \ \  P_{site}=p \, P_{k,c} \ . 
\eeq
For all $k>2$ a solution with non-zero $P$ is found at large values of $p$; the solution disappears at a critical value $p=p_c$ with a square-root singularity thus exhibiting the celebrated mixed first-order/second-order character.

As shown in the Supplemental Material for \cite{rizzo2019fate} the connected probability that two points at large distances $L$ on the Bethe lattice are both on the $k$-core is given:
\beq
c_L(p) \approx {p_1^2\, \Delta \over \mu}\,L\, \mu^{L}+O(\lambda^L)\ 
\eeq
where 
\beq
\mu=p {c-2 \choose k-2 }P^{k-2}(1-P)^{c-k} 
\eeq
Note that in \cite{rizzo2019fate} the parameter $\mu$ is called $\lambda$ but we changed notation to avoid confusion with the MCT parameter exponent.
\beq
\Delta \equiv {1-\mu \over \mu} P(1-P) \ 
\eeq
\beq
p_1 \equiv p {c-1 \choose k-1} P^{k-1} (1-P)^{c-k}\ 
\eeq   
In the case $(c=4,k=3)$ we have 
\beq
P={3 \over 4}\, ,\  \mu={1 \over 3}\, ,\ \ P_{site}={21 \over 32} \, \ \mathrm{for} \ p=p_c={8\over 9} \rightarrow T_c={1 \over \ln 8} \approx .480898
\eeq
From which the expressions for $c_1$ and $c_2$ given in section below can be obtained.

\section{Computing the parameters of SBR}
\label{sectionCC}

In this section we will compute the parameters $\alpha$, $\Delta \sigma$ and $\sigma$ of SBR for FA models on $M$-layer lattices by comparing the equal-time fluctuations of the density $g(x,t)$ in the Mean-Field regime. 
According to the analysis of \cite{altieri2017loop} on the $M$-layer lattice the leading (mean-field) expression of the correlation of the order parameter at two points of the lattice is given by the correlation on all non-backtracking paths connecting the two sites on the original lattice divided by $M$:
\beq
\langle g(x,t) g(y,t) \rangle- \langle g(x,t) \rangle \langle g(y,t) \rangle={1 \over M}\sum_{L=0}^{\infty} N_L(x,y) c_{L}^{Bethe}(t)
\label{2pM}
\eeq
where $c_{L}^{Bethe}(t)$ is the correlation of the shifted persistence between two sites $i$ and $j$ at distance $L$ on the Bethe lattice:
\beq
c_{L}^{Bethe}(t) \equiv \langle g_i(t) g_j(t) \rangle- \langle g_i(t) \rangle \langle g_j(t) \rangle
\eeq
and $N_L(x,y)$ is the number of non-backtracking walks between point $x$ and point $y$ on the original lattice (corresponding to $M=1$).
As we have discussed in the main text, for $M$ finite but large, in the regime where SBR  provides a quantitatively accurate description, the length-scale of the fluctuations of $g(x,t)$  is large and thus we are interested in the regime where $x-y$ (and thus $L$) in $N_L(x,y)$ is also large. In this regime $N_L(x,y)$ tends to a Gaussian with a $O(L)$ variance:
\beq
N_L(x,y) \
\approx {c(c-1)^{L-1} \over \rho}G(x-y)\ .
\eeq 
The above relationship is valid only if $x$ and $y$ corresponds to coordinates of  points of the lattice and is zero otherwise. This explains the prefactor, indeed since as we will see the variance of the Gaussian is much larger than the lattice spacing we can replace the sum over lattice points as an integral on the continuum  $\sum_j \rightarrow \int \rho\, d^dx)$. The integral on the other hand must be equal  
 to the total number of paths of length $L$ originating from a point, {\it i.e.} $c (c-1)^{L-1}$.
In terms of the unitary Fourier transform we can then write
\beq
N_L(k,k') \approx \delta(k+k')  {c \over \rho (c-1)} (c-1)^L  \, \exp[-L\, D_{NBW} \, k^2 ]
\label{Nkk}
\eeq 
Where $D_{NBW}$ is by definition the diffusion coefficient of non-backtracking random walks $\{\mathbf{x}_1,\dots, \mathbf{x}_t\}$ on the original lattice with $M=1$:
\beq
D_{NBW}\equiv \lim_{t \rightarrow \infty}{\langle ||{\bf x}_t||^2 \rangle \over 2 \, d\, t }  \ .
\eeq
In section \ref{sec:diffcoe} we will provide a simple expression for $D_{NBW}$ in terms of the dimension of the lattice and of its connectivity valid for a huge class of lattices.

If we compute quantities at the level of the MF approximation we will get the spurious dynamical arrest transition. In particular, the FA model on the Bethe lattice below the critical temperature (corresponding to $p<p_c=8/9$) displays a glassy phase. In the glassy phase it is convenient to study the persistence and its fluctuations in the infinite-time limit. Furthermore in this limit the critical properties of the blocked (negative) sites are exactly the same of the $k$-core of bootstrap percolation.
In particular we have
\beq
\lim_{t \rightarrow \infty}\langle g(x,t) \rangle=  P_{site}(p)-P_{site}(p_c)\approx {27 \over 16 \sqrt{2}} \delta p^{1/2}
\label{gMl}
\eeq 
and 
\beq
\lim_{t \rightarrow \infty}c_{L}^{Bethe}(t)=c_L^{Bethe}(p) \approx c_1 \, L \, \mu^{L}(p)
\label{clBethe}
\eeq 
where $P_{site}(p)$ is the $k$-core density, $\delta p \equiv p-p_c$ and $c_L^{Bethe}(p)$ is the two-point correlation of the $k$-core. The above formula for $c_L^{Bethe}(p)$ (first appeared in \cite{schwarz2006onset}) is derived in the supplemental material of Ref. \cite{rizzo2019fate} (see also section \ref{sec:conn} above) where it is also shown that  $\mu(p)$  tends to the critical value $(c-1)^{-1}$ at large times with a correction of order $\sqrt{\delta p}$
\beq
\mu(p) \approx {1 \over c-1}\, (1-c_2 \,  \delta p^{1/2}) 
\eeq
and the numerical constants $c_1$ and $c_2$ read for $c=4$ and $m=2$ \cite{rizzo2019fate}:
\beq
c_1 = {81 \over 512} \, , \ \  c_2={3 \over \sqrt{2}}
\eeq
In section \ref{sec:conn} we wrote the general formulas from which $c_1$ and $c_2$ were obtained.
Putting eqs. (\ref{2pM}), (\ref{Nkk}) and (\ref{clBethe}) together and performing the summation over $L$ we can write for the unitary Fourier transform of the fluctuations:
\beqd
\lim_{t \rightarrow \infty} (\langle g(k,t) g(k',t) \rangle- \langle g(k,t) \rangle \langle g(k',t) \rangle) = 
\eeqd
\beq
 = {1 \over M} \delta(k+k')  {c_1  \, c \over  \rho \, (c-1) }\left({1 \over c_2 \, \delta p^{1/2}+ D_{NBW} \, k^2 }\right)^2 \ .
\label{ggMl}
\eeq
We stress the difference from a simple Ornestein-Zernicke form due to the square that appears because of the $O(L)$ prefactor in (\ref{clBethe}).
The above expression has to be compared with the long-time limit of the mean-field approximation to SBR in the glassy phase $\sigma > 0$. The MF approximation to SBR corresponds to assume $\Delta \sigma^2$ is negligible and leads to:
\beq
\lim_{t \rightarrow \infty}[ \hat{g}(x,t) ]= \left( {\sigma \over 1-\lambda}\right)^{1/2} \ .
\label{gSBR}
\eeq 
The MF approximation to fluctuations can be computed treating the fields $h(x)$ as a small perturbation and is given by:
\beqd
\lim_{t \rightarrow \infty}([ \hat{g}(k,t) \hat{g}(k',t) ]- [ \hat{g}(k,t) ][ \hat{g}(k',t) ])= 
\eeqd
\beq
= \Delta \sigma^2 \delta(k+k')  
\left({1 \over 2\sqrt{(1-\lambda) \sigma}+ \alpha \, k^2 }\right)^2
\label{ggSBR}
\eeq
By equating expressions (\ref{gSBR}) and (\ref{gMl}) we obtain:
\beq
\sigma=(1-\lambda){729 \over 512}(p-p_c)
\eeq
By equating expressions (\ref{ggSBR}) and (\ref{ggMl}) we obtain:
\beq
\Delta \sigma^2={1 \over M \, \rho} \, {2187 \over 8192}(1-\lambda)^2 
\label{Dsigma}
\eeq
\beq
\alpha=(1-\lambda){9 \over 8}\, D_{NBW} 
\eeq
Thus the microscopic properties of the original lattice ($M=1$) enter through the density of lattice points $\rho \equiv N/V$ and through the diffusion coefficient on non-backtracking walks $D_{NBW}$. For the diamond lattice we have $\rho=1/8$ and $D_{NBW}=1$ as can be obtained from a general formula derived in  section \ref{sec:diffcoe}.
Replacing $\rho=1/8$, $D_{NBW}=1$ and $\lambda=.634$ (from the numerical solution on the Bethe lattice discussed before) we have obtained the values quoted in the main text.

Note that although we have considered the three-dimensional diamond lattice the previous relationships are valid for a generic lattice of connectivity four and can be used to repeat the numerical analysis on many other lattices, including notably the regular two-dimensional lattice where $\rho=1$ and $D_{NBW}=1/2$. Formulas for lattices with generic connectivity and topology can be obtained from section \ref{MFbehavior} and \ref{sec:diffcoe}.

Note that the MF approximations we have discussed are utterly wrong because neither the FA model on the $M$-layer nor the SBR equation actually display  a dynamic arrest transition and there is no static limit. Nevertheless we are making in both cases the same (wrong) approximation and it is thus correct to compare the outcome.

\section{Dynamical Mean-Field Fluctuations}
\label{sectionT}

In the previous section we have computed the mean-field expression (\ref{2pM}) of the fluctuations in the glassy phase and in the infinite time limit. In the following we will obtain its expression at large times at the critical point ($p=p_c=8/9$). In this case one expects (and may confirm numerically) that:  
\beq
c_L^{Bethe}(t) \approx c_1 \, L \, \mu^{L}(t)
\eeq
where $\mu(t)$ tends to the critical value $(c-1)^{-1}$ at large times with a correction of order $t^{-a}$
\beq
\mu(t) \approx {1 \over c-1}\, (1-\tilde{c}_2 \, t^{-a})
\eeq
The  constants $c_1$ is the same obtained above in the glassy phase, while $\tilde{c}_2$ is different from $c_2$ and at present can only be extracted from the numerics.   
Repeating the same steps above we then obtain:
\beq
\langle g(k,t) g(k',t) \rangle- \langle g(k,t) \rangle \langle g(k',t) \rangle
={1 \over M} \delta(k+k') {c_1 \, c \over \rho \,(c-1) }\left({1 \over \tilde{c}_2 t^{-a}+ D_{NBW}\, k^2 }\right)^2
\label{x4ML}
\eeq
this expression has to be compared with the mean-field small-time approximation of the SBR equations.
This is obtained  solving the equations perturbatively around the small-time limit $\hat{g}(x,t) \approx (t/t_0)^{-a}$. In Fourier space one computes the correction  due to the $s(x)$, then  after averaging one easily obtains:
\beq
[ \hat{g}(k,t) \hat{g}(k',t) ]- [ \hat{g}(k,t) ][  \hat{g}(k',t) ]
= \Delta \sigma^2 \delta(k+k')  
\left({1 \over  \, A_1^{-1} (t/t_0)^{-a}+  \alpha \, k^2 }\right)^2
\label{x4SBR}
\eeq
where we have used Gotze's definition (eq. 6.63a in \cite{gotze2008complex}) 
\beq
A_1 \equiv {1 \over 2  (a\, \pi \, \csc (a \pi)-\lambda)}\ .
\eeq
Note the above expression will be modified at larger times and should be replaced with the full SBR average.
Equating (\ref{x4ML}) and (\ref{x4SBR}) provides an alternative way to determine the SBR parameters $\Delta \sigma$ and $\alpha$ from the dynamics of FA models on the Bethe lattice (through the constants $c_1$ and $\tilde{c}_2$).

Setting for simplicity $\Delta \sigma^2=1/M$, $\alpha=t_0=1$ we obtain  
\beq
\langle g(k,t) g(k',t) \rangle- \langle g(k,t) \rangle \langle g(k',t) \rangle
= {t^{2 a} \over M} \delta(k+k')  
\left({1 \over 1+  (k \, \xi(t) )^2 }\right)^2\, , \ \ \ \xi(t) \propto t^{a/2} \,,\  t \gg 1 
\label{ggdyn}
\eeq
Thus at the MF level the total susceptibility diverges with time as  $t^{2a}$ and the correlation length diverges as $t^{a/2}$. Both these behaviors will change when the system starts to deviate from MF: the correlation length will not increase indefinitely and $g(x,t)$ will become negative at a finite time.
In this regime corrections of all orders in $1/M$ become equally relevant and one must abandon MF theory, technically using SBR in place of the MF espressions  eq. (\ref{ggdyn}) amounts to include corrections at {\it all} orders in powers of $1/M$ \cite{rizzo2016dynamical}.
We note {\it en passant} that MF fluctuations do not have the Ornstein-Zernicke form but rather the square of it.
Thus it should be noted that, while the use of the OZ form to fit numerical data of dynamical fluctuations is widespread in the literature \cite{lavcevic2003spatially,berthier2007spontaneous,
flenner2013dynamic,karmakar2014growing},  it has no theoretical justification. Both MF theory and SBR certainly do not support OZ in the $\beta$ regime, while the $\alpha$ regime predictions are at present not available, not even in MF theory. 
In real space the above expression leads to:
\beq
\langle g(x,t) g(y,t) \rangle- \langle g(x,t) \rangle \langle g(y,t) \rangle= {t^{a (2 -d /2)} \over M}  f\left( {x-y \over \xi(t)}\right)\ .
\eeq
that has been used in the main text to discuss the time-dependent Ginzburg criterion.
From the above expressions we also obtain the MF behavior of the $\chi_4(t)$ susceptibility at large times is
\beq
\chi_4(t) \approx M \, \rho\, \Delta \, \sigma^2 A_1^2  (t/t_0)^{2\, a} =  \, {2187 \over 8192}(1-\lambda)^2  A_1^2 \, (t/t_0)^{2 \, a} \approx  {2187 \over 8192}(1-\lambda)^2  A_1^2 \,g^{-2}
\eeq 
that gives the coefficient of the MF asymptote discussed in the main text.

\section{The SBR equations in universal form}
\label{sec:rescaling}

The equations of SBR are \cite{rizzo2014long}: 
\beq
\sigma  + s(x) =- \alpha \nabla^2 \, \hat{g}(x,t)-\lambda \, \hat{g}^2(x,t)+{d \over dt}\int_0^t \hat{g}(x,t-s)\hat{g}(x,s)ds
\eeq
where the field $s(x)$ is a {\it time-independent} random fluctuation of the separation parameter, Gaussian and delta-correlated in space:  
\beq
[s(x)]=0\, ,\ [s(x)s(y)]=\Delta \sigma^2  \, \delta (x-y) 
\eeq
and they have to be solved with the condition
\beq
\lim_{t \rightarrow 0} \hat{g}(x,t) (t/t_0)^a=1 \ .
\eeq
Thus the equations depend on five coupling constants: $\lambda$, $t_0$, $\alpha$, $\sigma$ and $\Delta \sigma$. 
However one can fix $t_0=\alpha=\Delta \sigma=1$  and then the general solution can be obtained by rescalings. This means that in practice the SBR equations at fixed $\lambda$ need only to be solved for varying values of $\sigma$.
More precisely one can easily verify that for a generic $K$-point function we have
\beq
[ \hat{g}(x_1,t_1) \dots \hat{g}(x_K,t_K) ]_{\alpha,\Delta \sigma,t_0,\sigma,\lambda}=
b_\phi^K [ \hat{g}(x_1/b_x,t_1/b_t) \dots \hat{g}(x_K/b_x,t_K/b_t) ]_{1,1,1,\sigma/b_\sigma,\lambda}
\eeq
where the notation $[\dots]_{\alpha,\Delta \sigma,t_0,\sigma,\lambda}$ means  that the SBR equations above are to be solved with the corresponding values of the five parameters. The rescaling parameters read:
\beqa
b_\phi & = & \Delta \sigma^{4 \over 8-d}\alpha^{-{d \over 8-d}}
\\
b_{x} & = & (\alpha/b_\phi)^{1/2}
\\
b_\sigma & =b_\phi^2
\\
b_t & = & t_0\, b_\phi^{-1/a}\ .
\eeqa

\section{The interplay between $M$ and the system size $L$}

In the main text we saw that in the MF regime valid at initial times the correlation length grows with time as $t^{a/2}$ while the time where SBR replaces MF theory grows with $M$ as $O(M^{{1 \over a( 4-d/2)}})$. Thus if the correlation length becomes comparable with the system size {\it before} deviations from MF occur the SBR equations do not have significant spatial variation and can be replaced by the SBR equations without the gradient term that will be written below. 
{\it This situation will always occur if we increase $M$ keeping the size $L$ of the original ($M=1$) lattice fixed}. 
This can be seen starting from the fact (see section {\ref{sec:rescaling}) that the solution of SBR corresponding to generic values of $\alpha$, $\Delta \sigma$ and $t_0$ can be expressed in terms of the solution with $\alpha=\Delta \sigma=t_0=1$ through appropriate rescalings of the correlators ($g \rightarrow b_\phi \, g$), of distances ($x \rightarrow b_x \, x$) and of times ($t \rightarrow b_t \, t$).

From the corresponding formulas we get that the $M$-dependence of the rescaling factors is the following:
\beq
\Delta \sigma^2 = O(M^{-1}) \rightarrow 
\left\{\begin{array}{ccc}
b_\phi & = & O(M^{-{1 \over 4-d/2}})
\\
b_x & = & O(M^{{1 \over 8-d}})
\\ b_t & = & O(M^{{1 \over a( 4-d/2)}})
\end{array}\right. 
\label{scalings}
\eeq
for generic dimension $d$ smaller than the critical dimension $d_c=8$.
The above relationships allow to understand the phenomenology on the $M$-layer lattice at the avoided MCT transition. They tell us that at $T_c$ deviations from the Bethe lattice behavior occur at times that increase with $M$ as $O(M^{{1 \over a( 4-d/2)}})$. 
Furthermore at times $O(M^{{1 \over a( 4-d/2)}})$ the order parameter, {\it i.e.} the deviations of the persistence from the plateau value, is small $O(M^{-{1 \over 4-d/2}})$ and fluctuates over a large length-scale $O(M^{{1 \over 8-d}})$, thus justifying {\it a posteriori} the validity of an effective theory description and the expectation that SBR predictions becomes increasingly accurate for larger values of $M$. In the main text we have given instead an {\it a priori} justification, starting from the mean-field expression of fluctuations in the $M$-layer and applying a time-dependent Ginzburg criterion that also lead to the above scaling eq. (\ref{scalings}).

According to the previous formulas the system size $L$ corresponds to an {\it effective} system size $L/b_x= L\, M^{-{1 \over 8-d}}$ in the universal theory.
Thus if we  if we go from $M$ to a larger value $M'$ while keeping the system size $L$ fixed we are systematically {\it reducing} the effective system size until we have a system so small that 
 spatial fluctuations of $g(x,t)$ are negligible because  the gradient term suppresses fluctuations on small distances.
In this regime the only source of fluctuations is the fluctuations of $\sigma$. The system becomes effectively a zero-dimensional system in the sense that the predictions of the finite-dimensional SBR equations are indistinguishable from those of the the zero-dimensional SBR equations discussed below, eqs. (\ref{SBRequazero}). 

To complete the discussion on the interplay between $M$ and $L$  it is interesting to consider what happens while changing $L$ at fixed large $M$. When $L=\infty$ we are in the thermodynamic limit and this is so also in terms of the effective system size $L/b_x$. 
For $L<\infty$ finite-size effects will be significant as soon as $L$ is of the order of the correlation length.
In this regime the solution of the SBR equations depends on the size  of the system and it is important to solve them  
in a box of the correct size to get accurate results.
If we decrease the system size further we will eventually reach the zero-dimensional limit: the spatial fluctuations of the solution are so small  that the finite-dimensional SBR equation can be replaced by the zero-dimensional equation discussed below for all practical purposes. 
 
The data for $L=4$ and values of $M=50000$, $M=100000$, $M=200000$ are actually in the zero-dimensional regime and the corresponding SBR predictions plotted of fig. 1 in the main text were obtained indeed from the zero-dimensional SBR equation. The data for $M=3000$ were obtained with $L=8$ an are instead strictly three-dimensional because, although the system size is finite, we had to consider the full three-dimensional SBR equation that depends on the value of $D_{NBW}$ on the lattice and on the system size.

\section{Finite-Size Corrections to Mean-Field Models (SBR in Zero Dimension)} 
\label{zerodim}
 
The zero-dimensional equations can be written solely in terms of the total shifted persistence
\beq
g(t)={1 \over M L^3 \rho}\sum_{i,\alpha} g_i^\alpha(t)\ .
\eeq
Then we have for a generic cumulant of order $K$:
\beq
\langle g^K(t)\rangle \approx [\hat{g}^K(t)]\, , \ \  1\ll M < \infty
\label{zeroequivalence}
\eeq
where in the RHS  $g(t)$ is the solution of of the zero-dimensional SBR equations \cite{rizzo2014long,rizzo2016dynamical,rizzo2015qualitative}:
\beq
\sigma  + s =-\lambda \, \hat{g}^2(t)+{d \over dt}\int_0^t \hat{g}(t-s)\hat{g}(s)ds \ .
\label{SBRequazero}
\eeq
The square brackets mean average with respect to  the field $s$ that is a {\it time-independent} Gaussian random fluctuation of the separation parameter:  
\beq
[s]=0\, ,\ [s^2]=  \Delta {\hat \sigma}^2  \ .
\eeq
The equations have to be solved with the small-time condition
\beq
\lim_{t \rightarrow 0} \hat{g}(t) (t/t_0)^a=1\ \ .
\eeq
Note that the gradient term is not present and indeed the actual geometrical structure of the lattice is irrelevant, the only relevant control parameter for SBR is the total number $N_{tot}= M \, L^3 \, \rho$ of sites of the lattice that controls $\Delta \sigma^2$ through: 
\beq
\Delta \hat{\sigma}^2={1 \over N_{tot}} \, {2187 \over 8192}(1-\lambda)^2 \, ,
\eeq
that follows from expression (\ref{Dsigma}). The above expression does not depend on how the lattice is actually generated and in particular it is also correct for a random-regular-graph of size $N_{tot}$ \cite{franz2013finite}.
In the present setting $N_{tot}=M\, L^3 \, \rho=8\, M$.

\begin{figure}[t]
\centering
\includegraphics[scale=.7]{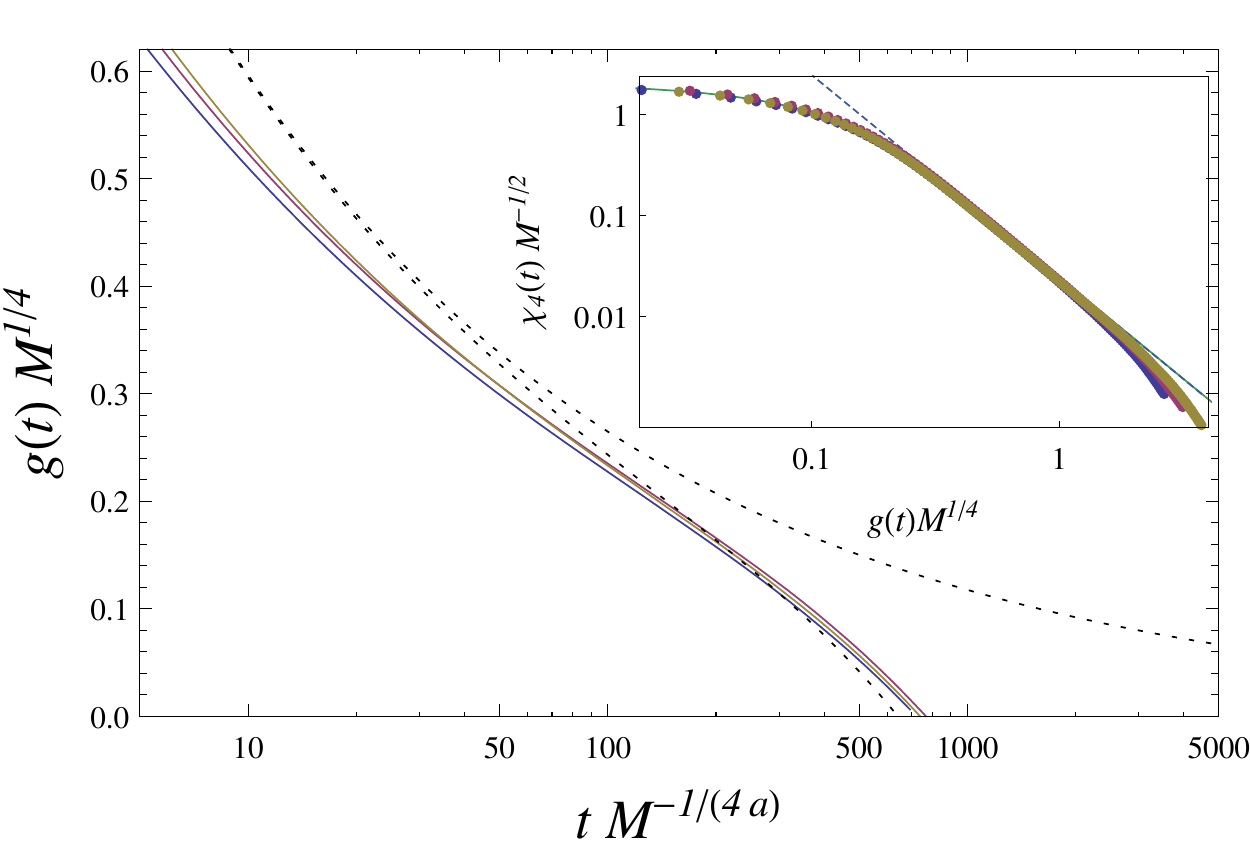} 
\caption{Rescaled shifted persistence for $L=4$ and $M=5 \times 10^4$, $M=10^5$ and $M=2\times 10^5$. 
The dotted lines  are the mean-field asymptotic expression $(t/t_0)^{-a}$ (top) and the SBR prediction (bottom). 
The data for different $M$ display a good collapse on the SBR curve at large rescaled times. Inset: scaling parametric plot of $\chi_4(t)$ vs. $g(t)$ data for various $M$ (points) collapse onto a single SBR curve (solid).}
\label{fig:ptd0res}
\end{figure}

As we said mentioned in previous section the data for $L=4$ and values of $M=50000$, $M=100000$, $M=200000$ are in the zero dimensional regime.
In fig. (\ref{fig:ptd0res}) we rescaled them  horizontally and vertically with the appropriate powers of $M$ that  lead to collapse  on a unique SBR curve. The rescaling factors are $M^{1/4}$ for the  shifted persistence and $M^{-1/(4 a)}$ for time and are obtained setting $d=0$ in eqs. (\ref{scalings}), which is the technical reason why finite-size corrections correspond to zero dimension. 

Note that the agreement between data and SBR in the scaling plot of the figure is not yet perfect and this can be tracked to the fact that, at the values of $M$ considered, the curves start to deviate from the mean-field curve at times where the $M=\infty$ mean-field curve itself has still relevant small-time corrections of order $(t/t_0)^{-2 a}$ to the leading asymptotic behavior $(t/t_0)^{-a}$, as discussed in section \ref{MFbehavior}. As a consequence the data for small rescaled times $t M^{1/(4a)}$ of order  $5-20$ display an approximate $20\%$ deviation from the asymptotic expression $(t/t_0)^{-a}$ that should describe the data for {\it small} rescaled times at large values of $M$ but the agreement improves considering larger values of $M$.
On the other hand small-time corrections gets smaller at larger rescaled times $\approx 50$ and correspondingly the data agree much better with the SBR curve. 
We note that the rescaling factor $M^{1/4}$ that leads to the collapse of the curves was obtained for the first time in \cite{franz2011field} from a static computation, but the resulting scaling function, apart from being independent of time, is ill-defined and one needs a full-fledged dynamical treatment to compute the time-dependent scaling curve.

\section{Diffusion Coefficient of Non-Backtracking random walks on generic lattices}
\label{sec:diffcoe}

We consider lattices with connectivity $c$ such that we can classify sites in two classes, say red and blue, such that a blue  site is connected to the $c$ red  sites located in the $c$ directions ${\mathbf v}_\mu$.
Similarly a red site is connected to the $c$ blue sites located in the $c$ directions $-{\mathbf v}_\mu$.
We will {\it not} require the condition $\{{\mathbf v}_\mu\}=-\{{\mathbf v}_\mu\}$, meaning that the set of directions need not to be invariant under inversion of the coordinate axes. Thus we extend previous results \cite{altieri2017loop,fitzner2013non}  to include {\it e.g.}  the honeycomb lattice in $d=2$ and the diamond lattice in $d=3$.
We assume instead that
\beq
\sum_\mu {\bf v}_\mu=0\, ,\ \ ||{\bf v_\mu}||=v \, .
\eeq 
A non-backtracking random walk can be described as a sequence  of steps at consecutive times $s$ in the directions $\mu(s)$:
\beq
{\bf x}_t=\sum_{s=0}^{t-1}(-1)^s {\bf v}_{\mu(s)}
\eeq
where the minus sign comes from the fact that  the set of possible directions changes as the walker moves from a blue and to a red site.
The average can then be written in terms of the joint probability $P^{(s,s')}_{\mu,\mu'}$
\beq
\langle ||{\bf x}_t||^2 \rangle = \sum_{s,s'=0}^{t-1}(-1)^{s+s'} {\bf v}_{\mu} \cdot {\bf v}_{\mu'} P^{(s,s')}_{\mu,\mu'} \ ,
\eeq
note that we use Einstein's convention of implicit summation over repeated indexes $\mu$.
Since the directions are uniformly distributed at zero time we have exactly
\beq
P^{(s,s')}_{\mu,\mu'} = 
P^{(0,s-s')}_{\mu,\mu'} 
\eeq
and we can write for large $t$
\beq
\langle ||{\bf x}_t||^2 \rangle \approx t \, v^2 \left(2 \sum_{s=1}^{\infty}(-1)^{s} {\bf\hat{v}}_{\mu} \cdot {\bf \hat{v}}_{\mu'} P^{(0,s)}_{\mu,\mu'} +1\right)
\label{larget}
\eeq
The probability can be computed recursively
\beq
P^{(0,s+1)}_{\mu,\mu'}=T_{\mu,\mu''}P^{(0,s)}_{\mu',\mu''}
\eeq
where $T_{\mu,\mu''}$ is the $c\times c$ matrix with zero diagonal elements and off-diagonal elements equal to $1/(c-1)$. We can thus write 
\beq
T={\mathcal P}_s-{1 \over c-1}{\mathcal Q}_s
\eeq
where ${\mathcal P}_s$ is the projector on the vector with all equal components and ${\mathcal Q}_s= I-{\mathcal P}_s$ is the orthogonal projector. It follows that 
\beq
P^{(0,0)}= {1 \over c} \, I\, , \ \ 
P^{(0,s)}= {1 \over c} T^s=  {1 \over c} {\mathcal P}_s+{1 \over c}\left({-1 \over c-1}\right)^s{\mathcal Q}_s
\eeq 
since ${\mathcal P}_s$ is a matrix with all elements equal it gives zero contribution when summed over the directions because of the condition $\sum_\mu {\bf \hat{v}}_\mu=0$ and we have
\beq
{\bf \hat{v}}_{\mu} \cdot {\bf \hat{v}}_{\mu'} P^{(0,s)}_{\mu,\mu'}=\left({-1 \over c-1}\right)^s v^2
\eeq
replacing the above expression in the large time expression eq. (\ref{larget}) we obtain:
\beq
\langle ||{\bf x}_t||^2 \rangle \approx  t \, {c \, v^2 \over c-2}
\eeq
and therefore (taking into account that each vector has $d$ components)
\beq
D_{NBW} \equiv \lim_{t \rightarrow \infty}{\langle ||{\bf x}_t||^2 \rangle \over 2 \, d\, t }  = {c  \, v^2 \over 2 \, d \, (c-2)}\ .
\eeq
On regular lattice $c=2 d$ and we recover the result $D_{NBW}=1/(2\,d-2)$. On the diamond cubic lattice studied in the paper we have $c=4$ and $v^2=3$ leading to $D_{NBW}=1$. 
The honeycomb lattice can be realized repeating a  $3 \times \sqrt{3}$ unit cell with two blue points at coordinates $(0,0)$,$(\sqrt{3}/2,3/2)$ and two red points with coordinates $(0,1)$, $(\sqrt{3}/2,5/2)$. This leads to $\rho=4 \, 3^{-3/2}$ and $D_{NBW}=3/4$.

\end{document}